# Mobile Game User Research: The World as Your Lab?

**Jan Smeddinck**
University of Bremen
Bibliothekstraße 1
Bremen 28359, Germany
smeddinck@tzi.de

**Markus Krause**
University of Hannover
Research Group HCI
Hannover, Germany
markus.krause@
hci.uni-hannover.de

**Kolja Lubitz**
University of Bremen
Bibliothekstraße 1
Bremen 28359, Germany
pinguin@tzi.de

## Abstract
With the advent of mobile games and the according growing and competitive market, game user research can provide valuable insights and a competitive edge if methods and procedures are employed that match the distinct challenges that mobile devices, games and usage scenarios induce. We present a summary of parameters that frame the research setup and procedure, focusing on the trade-offs between lab and field studies and the related decision whether to pursue large-scale and quantitative or small-scale focused research accompanied by qualitative methods. We then illustrate the implications of these considerations on real world projects along the lines of two evaluations of different input methods for the action-puzzle mobile game Somyeol: a local study with 37 participants and a mixed design of qualitative and quantitative methods, and the strictly quantitative analysis of game-play data from 117,118 users. The findings underline the importance of small-scale evaluations prior to release.

## ACM Classification Keywords
K.8.0 [Personal Computing]: General - *Games*.

| **General Parameters:** |
|---|
| Lab or field, offline or online, qualitative or quantitative, focus (depth) or scale (breadth). |
| **Research Purpose:** |
| Understanding, engineering, re-engineering, evaluating, or describing (cf. [1]). |
| **Research Focus:** |
| Playability and/or player experience (with subcategories such as mobility, social aspects, etc.), hardware-inclusive or software-centric, academic or business, game-centric or fully situated. |
| **Interaction Modalities:** |
| Visual and/or sound and/or touch (cf. [2]). |

**Table 1:** Framing considerations for mobile HCI / mobile game user research planning.

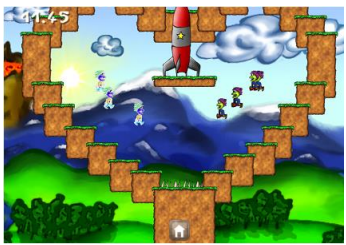

*Somyeol* is a puzzle platform game in which the player controls more than one character at a time and tries to steer as many of his/her Somyeols as possible through the levels.

## Introduction

In 2011, mobile games represented a worldwide market of $5.8 billion[1]. Mobile game user research (mGUR) is an important part of mobile game development. It can give developers a competitive edge and is a subject for scientific research given the close relation to mobile HCI. Despite of the market size, stable methods and procedures for mGUR are still sparse. While some methods that stem from general GUR can be used for mGUR, the mobile players' context differs from non-mobile situations. Gamers potentially change their location, or their environment while playing, or may play in public places where their potential to focus exclusively on a game is limited. Such factors must be considered in mGUR and mobile user research in general. For the meanwhile, the development and evaluation of mobile systems is often still based on trial and error instead of on user-centered design [7]. One reason for this might be the anticipated challenge in designing and conducting mGUR. In this work we list the most common basic framing considerations that can provide a structured approach to planning mGUR. We also provide a short overview of two evaluations of the mobile game *Somyeol*; a small-scale local field study with a mixture of quantitative and qualitative methods and a strictly quantitative large-scale analysis of game-play data, together with a discussion of the framing considerations for this setup.

## Challenges of Mobile Game User Research

Most methods and procedures from non-mobile playability and game experience evaluations can be useful tools for evaluations with mobile devices, given the right circumstances and adjustments. The general mobile HCI literature can also be a good reference, since it is more established than the comparatively young field of mGUR. Still, interaction research with mobile devices is challenging for a number of reasons. Zhang and Adipat [10] warn of complications with mobile research, highlighting the challenges of: mobile context, limited connectivity, small screen size, low display resolution, limited processing power, and limited data entry methods. However, while some of these challenges still apply, this study, dating back only to 2005 demonstrates how rapidly things are changing in the mobile world. Many of these problems are much less severe nowadays and methods that provide good rigs for quick-to-setup and affordable mobile studies are increasingly available (e.g. [9]). Given the speed of developments in mobile devices, operating systems and usage patterns, game user researchers are still challenged to keep up with adjusting their methods to the changing circumstances. Taking a step away from specialized methods and setups, basic considerations remain more stable and can help with making informed decisions about adequate research setups.

[1] Newzoo Trend Report Mobile Games, March 2012

## Basic Considerations for Planning Mobile Game User Research

Related literature, especially from the area of general mobile user research provides detailed overviews of available research methods and procedures [4][7][8]. However, a systematic overview of framing parameters (cf. *Table 1*) provides a good structure for examining more general considerations that frame the approach to setting up successful mGUR studies. There is ongoing debate in the literature about the pros and cons of the different sides of the dimensions of the design space for mGUR laid out by these considerations, especially concerning the general parameters. For example, while

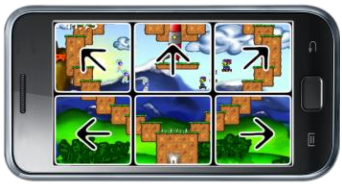

*MTG*: For the multi-touch grid method the screen is split into six areas. The grid maximizes button size. The lower left and right areas are used for walking, the top left and right areas for jumping in respective directions. The top center triggers a straight jump.

*ACC*: Uses the accelerometer to control left and right movements depending on the phone tilt. A touch gesture triggers jumping.

*OFR*: The one-finger relative input method uses the first touch point from the screen. If the touch point moves left, right, or up (while the player holds down his/her finger), the characters move accordingly.

some research suggests that there are no significant differences in the results from a lab- and a field-version of an otherwise unchanged mobile applications usability testing [6] (which would arguably also help to avoid challenges with field studies), others argue that significant differences exist [3] and point out that recent developments have made tests "in-the-wild" much more practical [8]. In this light, combinations of lab and field studies likely provide the most stable insights [2] and since artificial field studies do not necessarily result in natural settings, an augmentation with online studies that capture real player-data seems advisable [6] - especially since the rise in popularity of "appstores" facilitates widespread distribution [5].

Similar arguments can be found concerning the tightly connected balance between qualitative and quantitative research methods, as well taking as a deeply focused approach compared to aiming for broad scale. Since it may never be possible, even for carefully controlled comparative studies to weed out insecurities about the capabilities of the available user research methods to capture desired information on playability and game experience, regardless of game mechanics, approaching hypotheses with heterogeneously framed, triangulating research promises the most reliable results. The next section contains an illustration of such an approach with manageable complexity and cost.

## Ask 37 Players or 100,000?

*Somyeol*[2] is a mobile *Jump and Run* game. The player controls a group of characters which come in a number of different classes with different capabilities. All "*Somyeols*" are controlled at the same time with the same input. The goal is to safely steer them through an obstacle course. The game requires input for three actions: moving left/right and jumping. Determining adequate metaphors for these inputs was an important aspect of developing *Somyeol*. The implemented methods were *multi-touch grid* (MTG), *accelerometer* (ACC), *one-finger relative* (OFR), and *gamepad emulation* (GPE). GPE was only implemented for the release to market version (cf. sidebar for more details).

---

[2] http://somyeol.com/, last viewed 2013-01-18.

*A Study in the Tamed Wild*
In order to evaluate the game experience and playability of the metaphors, a small scale study was conducted. The study was carried out in the field in settings that constitute realistic environments for mobile gaming, such as train and tram stations, in public transport, or at the university. The 37 participants (10 f, 27 m) had and age group mode of 20–30 years (81% of all participants). They played three levels with three tries each. The first level only required moving the characters. The second and third level required moving and jumping. In the third level it was also possible to fail. The experiment was designed as a between-group study. After playing, participants completed a questionnaire which gathered general demographics, experience with mobile games, and their experience during game-play. Player performance data such as time-needed for a level, final scores, and the number of Somyeols that were lost was also logged. The quantitative results showed no consistent differences between the input modes. However, qualitative feedback and an analysis of the performance data suggested that the tested methods were not optimal and as a result, a fourth input metaphor (GPE, which more closely resembles gamepads) was added to the release version.

*Jan Smeddinck* is a doctoral candidate at the *University of Bremen, Germany* and holds a fellowship of the *Klaus Tschira Foundation* (KTS). Working in the fields of interaction design and serious games, he focuses on automated learning from interactions in adaptive systems.

*Markus Krause* is a research fellow at the human computer interaction research group at the *Leibnitz University*. He is engaged in research on interaction design, digital games, and human computation.

*Kolja Lubitz* is a master's student of the computer science study program at the *University Bremen*. He was an organizer of the Global Game Jam 2011 in Bremen and he is an independent game designer and developer at *Brain Connected*.

http://dm.tzi.de/en/people/staff/

## Acknowledgements
We would like to thank all members of the independent development team *Brain Connected*, all organizers of the *Global Game Jam 2011* in Bremen and the participants of the field study. This work was partially funded by the *Klaus Tschira Foundation* (KTS) through the graduate school *Advances in Digital Media*.

*Analyzing Data from the Real Wild*
A post-release comparative analysis was executed with data from 117,118 players (captured during 11 months in 2012). Only 3121 players changed the active input metaphor from the new default GPE to one of the other methods, which were available via the settings menu. These players changed the input method 13,285 times (M=4.26 times per player). 53% (1,663) reverted back to the GPE metaphor after trying other modes. 12% (370) last settled on *OFR*, 18% (571) on *MTG* and 17% (517) on *ACC*. Overall, the players showed a clear preference for GPE. All other modes were roughly en par (as suggested by the small-scale field study). All methods were similarly well considered before users made their final choice (percentage of sessions played while still making changes: *GPE*: 28%, *OFR:* 21%, *MTG*: 26%, *ACC*: 25% of a total of 136,680 sessions).

## Discussion
Our results illustrate the importance of a qualitative perspective, especially in evaluations during the design and implementation phase of mobile games. The quantitative analysis of a large number of game records helped solidify the decision for an alternative input mode and confirmed the equality of the other approaches. This allowed us to capture a more complete picture than an isolated study or a focus on one specific method would have provided. At the same time, the ad-hoc small scale field study did deliver valid insights. In order to inform the complex decisions that mGUR research faces, future comparative studies that include laboratory studies (to allow for more controlled testing and for employing research methods that are currently out of question for research in-the-wild) are needed to establish reliable connections between the framing considerations and specific research methods.

## References
[1] Abowd, G.D. and Mynatt, E.D. Charting past, present, and future research in ubiquitous computing. ACM Transactions on Computer-Human Interaction (TOCHI) 7, 1 (2000), 29–58.

[2] Baillie, L. and Schatz, R. Exploring multimodality in the laboratory and the field. Proceedings of the 7th international conference on Multimodal interfaces, ACM (2005), 100–107.

[3] Duh, H.B.-L., Tan, G.C.B., and Chen, V.H. Usability evaluation for mobile device: a comparison of laboratory and field tests. Proceedings of the 8th conference on Human-computer interaction with mobile devices and services, ACM (2006), 181–186.

[4] Hagen, P., Robertson, T., Kan, M., and Sadler, K. Emerging research methods for understanding mobile technology use. Proceedings of the 17th Australia conference on Computer-Human Interaction: Citizens Online: Considerations for Today and the Future, Computer-Human Interaction Special Interest Group (CHISIG) of Australia (2005), 1–10.

[5] Henze, N., Rukzio, E., and Boll, S. Observational and experimental investigation of typing behaviour using virtual keyboards for mobile devices. *Proceedings of the 2012 ACM annual conference on Human Factors in Computing Systems*, (2012), 2659–2668.

[6] Kallio, T. and Kaikkonen, A. Usability testing of mobile applications: A comparison between laboratory and field testing. Journal of Usability studies 1, 4-16 (2005), 23–28.

[7] Kjeldskov, J. and Graham, C. A Review of Mobile HCI Research Methods. In L. Chittaro, ed., Human-Computer Interaction with Mobile Devices and Services. Springer Berlin Heidelberg, 2003, 317–335.

[8] Kjeldskov, J. and Stage, J. New techniques for usability evaluation of mobile systems. International Journal of Human-Computer Studies 60, 5–6 (2004), 599–620.

[9] Schusteritsch, R., Wei, C.Y., and LaRosa, M. Towards the perfect infrastructure for usability testing on mobile devices. CHI '07 Extended Abstracts on Human Factors in Computing Systems, ACM (2007), 1839–1844.

[10] Zhang, D. and Adipat, B. Challenges, methodologies, and issues in the usability testing of mobile applications. International Journal of Human-Computer Interaction 18, 3 (2005), 293–30.